\documentclass{UApreprint}
\usepackage{graphics}
\usepackage{fullpage}
\usepackage{verbatim}
\usepackage[tight]{subfigure}
\usepackage{amssymb,amsmath}
\usepackage{citesort}
\newcommand{\be}{\begin{equation}}
\newcommand{\ee}{\end{equation}}
\newcommand{\bea}{\begin{eqnarray}}
\newcommand{\eea}{\end{eqnarray}}
\newcommand{\bean}{\begin{eqnarray*}}
\newcommand{\eean}{\end{eqnarray*}}
\newcommand{\bi}{\begin{itemize}}
\newcommand{\ei}{\end{itemize}}
\newcommand{\bdm}{\begin{displaymath}}
\newcommand{\edm}{\end{displaymath}}

\newcommand{\spc}{\hspace{.8mm}}
\newcommand{\Lam}{\Lambda_{\mathrm{QCD}}}
\newcommand{\ads}{\mathrm{AdS}_5}
\begin{document}

\thispagestyle{empty}
\vspace*{.5cm}
\noindent
\hspace*{\fill}{\large  OUTP-06 17 P}\\
\vspace*{2.0cm}

\begin{center}
{\Large\bf Aspects of Axion Phenomenology in a slice of $\boldsymbol\ads$}
\\[2.5cm]
{\large Thomas~Flacke and~David~Maybury}\\[.5cm]
{\it Rudolf Peierls Centre for Theoretical Physics\\
University of Oxford, 1 Keble Road, Oxford OX1 3NP, UK}
\\[.2cm]
(Dated: December 11, 2006)
\\[1.1cm]

{\bf Abstract}\end{center}
\noindent
Motivated by multi-throat considerations, we study the phenomenological implications of a bulk axion in a slice of $\ads$ with a large extra dimension: $k\sim 10^{-2}\, \mathrm{eV}$, $kR \gtrsim \mathcal{O}(1)$. In particular, we compare axion physics with a warped geometry to axions in flat compactifications. As in flat compactification scenarios, we find that the mass of the axion can become independent from the underlying Peccei-Quinn scale. Surprisingly, we find that in warped extra dimensions the axion's invisibility, cosmological viability, and basic phenomenology remain essentially unaltered in comparison to axions in flat compactifications.
\newpage

\setcounter{page}{1}

\section{Introduction}

The origin of CP violation within the standard model remains an essential theoretical problem. Current empirical evidence strongly favors the weak interactions as the source of observed CP violation through complex phases within the CKM matrix \cite{Eidelman:2004wy}. The lack of a Nambu-Goldstone boson interpretation for the $\eta$ meson, associated with the spontaneous breaking of a global U(1) flavour symmetry \cite{Weinberg:1975ui}, suggests the existence of a non-trivial QCD vacuum structure \cite{'tHooft:1976up}, which, in principle, violates CP. Extending the effective QCD Lagrangian to include the most general vacuum structure yields
\be
\mathcal{L}_{eff} = \mathcal{L}_{\mathrm{QCD}} + \bar \Theta \frac{g^2}{32\pi^2} F^{\mu\nu}_a \tilde F_{\mu\nu a}.
\label{QCD_1}
\ee
The effective strong CP violation parameter, $\bar\Theta$, contains contributions from both QCD and quantum flavour dynamics, namely
\be
\bar \Theta = \Theta + \mathrm{arg}\left(\mathrm{det}\spc M\right)
\ee
where $M$ denotes the quark mass matrix and $\Theta$ labels the CP violation term arising solely from QCD. From the low energy effective field theory point of view, $\bar\Theta$ appears as a standard model input. Empirical measurements, such as those arising from limits on the neutron's electric dipole moment indicate \cite{PDG},
\be
\bar \Theta \lesssim 10^{-9}.
\ee
A priori, $\bar \Theta$ could have a value anywhere on the interval $[0, 2\pi]$ and the origin of this parameter's seemingly unnatural tiny value constitutes the strong CP problem (see \cite{Peccei:1988ci,Peccei:2006as,Kim:1986ax} for reviews).

Perhaps the most elegant solution to the strong CP problem rests on the Peccei-Quinn (PQ) mechanism \cite{Peccei:1977hh}. In this scenario, $\bar \Theta$ becomes effectively promoted to a field -- the axion -- identified as a pseudo-Nambu-Goldstone mode associated with the spontaneous breaking of a global $\mathrm{U(1)}_{PQ}$ symmetry \cite{ww}. As the axion dynamically relaxes to the origin of its potential, $\bar\Theta$ vanishes. Augmenting the QCD Lagrangian of eq.(\ref{QCD_1}) by the PQ mechanism one finds,
\be
\mathcal{L}_{eff} = \mathcal{L}_{\mathrm{QCD}} + \frac{1}{2}\partial_\mu a \partial^\mu a + \frac{a}{f_{PQ}}\xi \frac{g^2}{32\pi^2} F^{\mu\nu a} \tilde F_{\mu\nu a}
\ee
where $f_{PQ}$ denotes the axion decay constant set by the scale of $\mathrm{U(1)}_{PQ}$ breaking, and the parameter $\xi$ arises as a model dependent factor. The axion obtains its mass  \cite{Bardeen_and_Tye} through QCD instanton effects that explicitly violate the $\mathrm{U(1)}_{PQ}$ symmetry and thereby provide the axion with a potential, yielding
\be
m_a \sim \frac{\Lam^2}{f_{PQ}}
\label{a_mass_4d}
\ee
where $\Lam \approx 250\spc\mathrm{MeV}$. Constraints from laboratory searches, astrophysics, and early universe cosmology bound $f_{PQ}$ (see \cite{Kim:1986ax,Raffelt} for reviews),
\be
10^{9}\spc\mathrm{GeV} \lesssim f_{PQ} \lesssim 10^{12}\spc \mathrm{GeV}
\ee
which, by eq.(\ref{a_mass_4d}), implies,
\be
10^{-5}\spc\mathrm{eV} \lesssim m_a \lesssim 10^{-2}\spc\mathrm{eV}.
\ee
We should emphasize that the PQ mechanism intimately links the scale of $\mathrm{U(1)}_{PQ}$ breaking, $f_{PQ}$, and the axion mass. The PQ mechanism and the resulting axion present a solution to the strong CP problem, however, the particulars of the UV embedding remain model dependent.\footnote{{\it Cf.} \cite{KSVZ,DFSZ} for the most well know implementations and see \cite{Kim:1986ax,Peccei:2006as} for reviews.}

One of the most striking axion scenarios results upon embedding the PQ framework within large extra dimensions \cite{DDGprecursors,DDG}. If the axion appears as a bulk pseudoscalar, novel features emerge \cite{DDG}, such as Kaluza-Klein (KK) mode interference, non-trivial axion mass relations, and increased energy dissipation among cosmological relic axions. With the advent of warped compactifications \cite{RS}, whereby large scale hierarchies can be understood through exponential red-shifting, several attempts were made in constructing warped axion models using both bottom-up approaches \cite{bottomup} and top down string constructions \cite{stringaxions,conlon}. In traditional warped extra-dimensional models that address the gauge hierarchy problem of the standard models, the first KK excitations of bulk fields occur near the electroweak scale, namely on the order of $10\,\mathrm{TeV}$. However, generic flux compactifications allow for backgrounds with several ``throats'' \cite{multithroats} originating from a compact Calabi-Yau manifold. As demonstrated in \cite{WA}, additional throats can provide new model building avenues for axions. In principle, the multiple throats could have widely different $\ads$ curvature radii and thus bulk fields would have their KK spectra dictated by the geometry of their throat. Motivated by these observations, we model-independently consider the extra-dimensional phenomenology of a bulk axion in a throat separate from the standard model with an AdS inverse curvature radius of $k \sim 10^{-2}\spc \mathrm{eV}$ and with $kR \sim \mathcal{O}(1)$. For simplicity, we will assume that the standard model remains confined to a UV brane. In this scenario, the axionic KK excitations appear hierarchically lower than the electroweak scale. The setup provides a warped extension to the flat extra-dimensional axion phenomenology studied in \cite{DDG}.

\section{Axions in a warped background}

\begin{figure}[ht!]
\newlength{\picwidthdo}
\setlength{\picwidthdo}{3.1in}
 \begin{center}
\subfigure[][]{\resizebox{\picwidthdo}{!}{\includegraphics{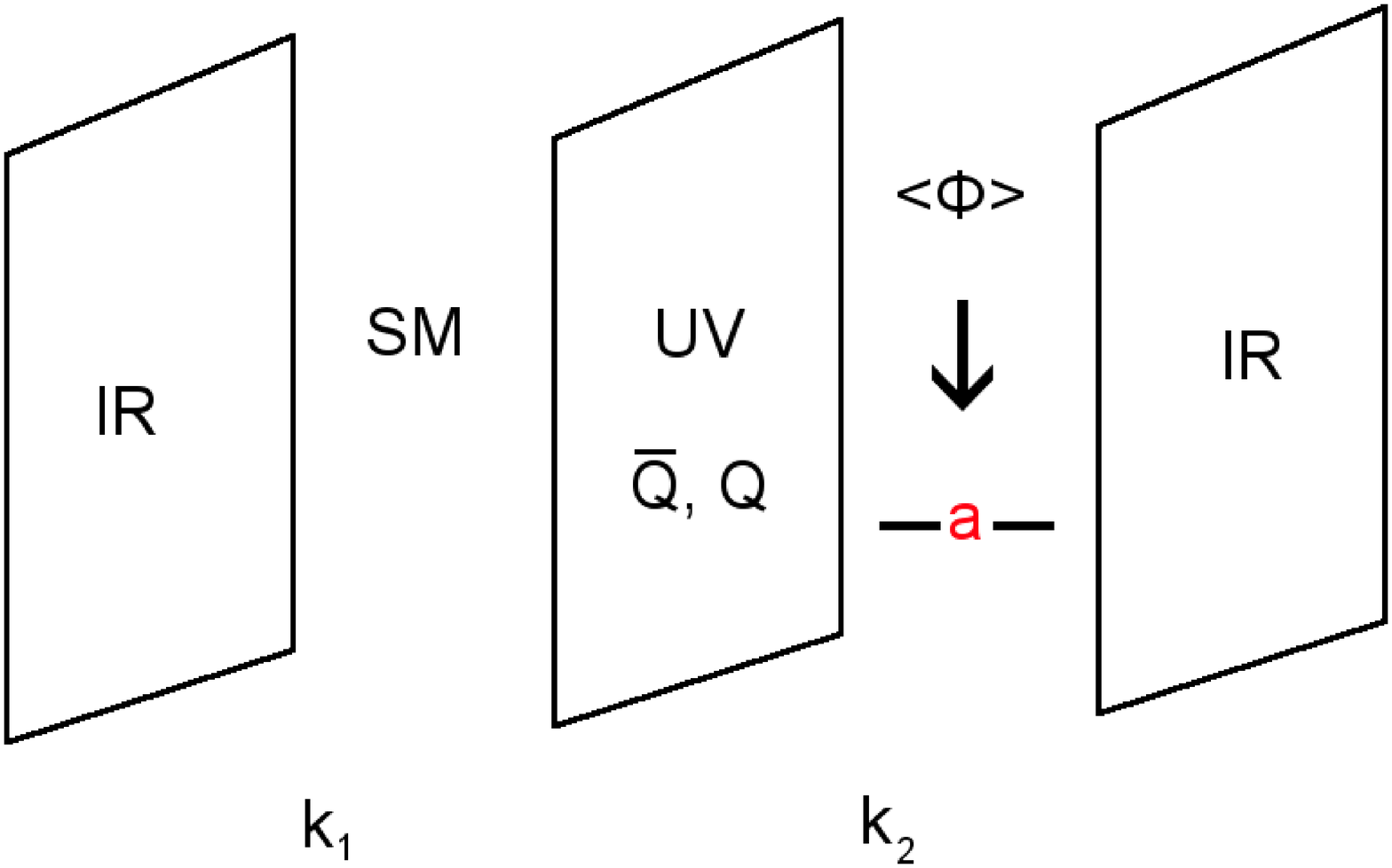}}}
\subfigure[][]{\resizebox{\picwidthdo}{!}{\includegraphics{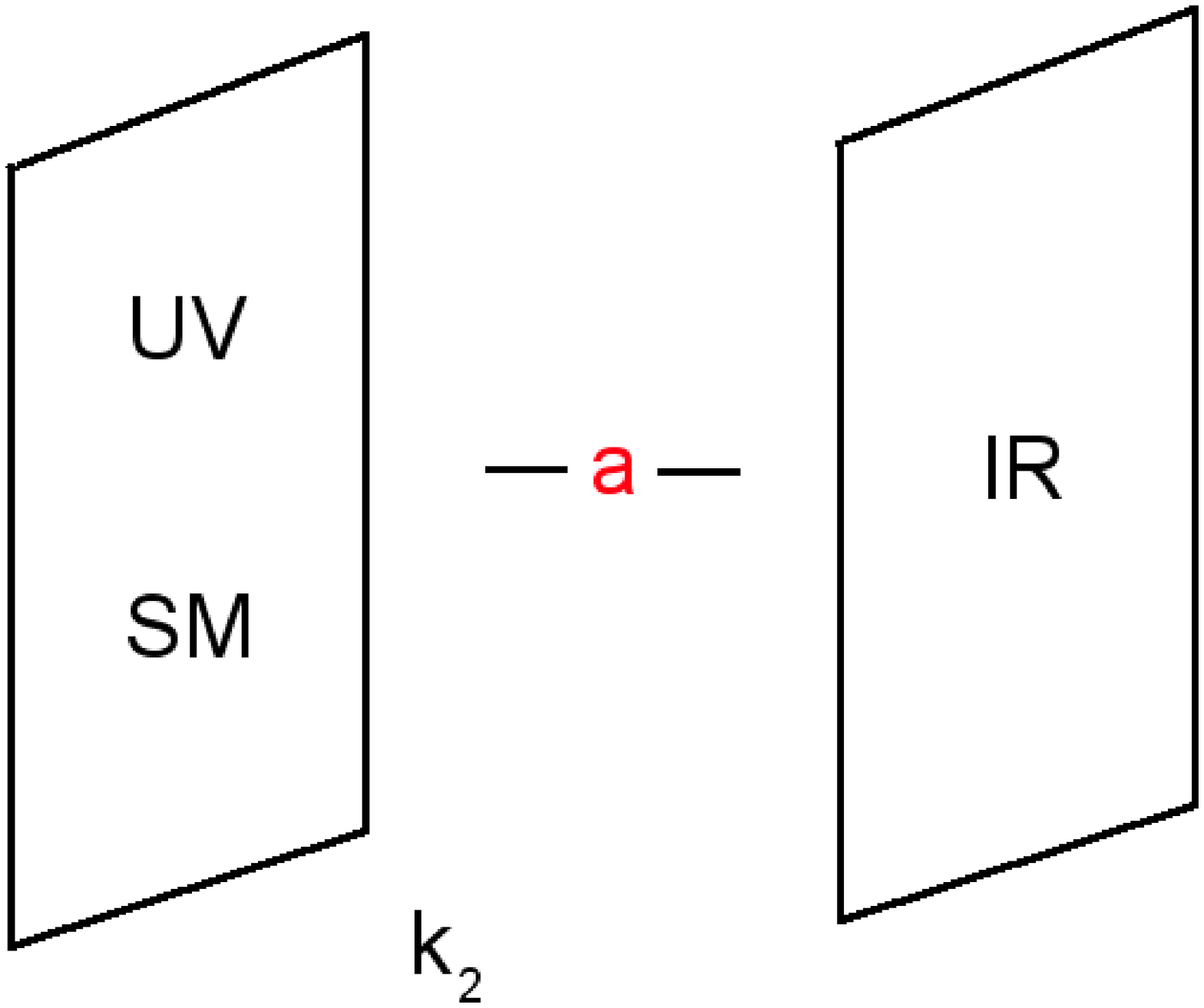}}}
%\subfigure[][]{\resizebox{\picwidthc}{!}{\includegraphics{kR_6_warped_wide.ps}}}
 \end{center}
\caption{Illustration of a two throat axion toy model. a) The standard model lives in left $\ads$ throat while a complex scalar field $\Phi$ with phase $a$ (the ``axion'') lives in the right throat. The two throats communicate via UV brane localized coloured fermions, $Q,\bar{Q}$. b) Once the PQ symmetry is broken, the resulting bulk pseudoscalar axion effectively sees the standard model as UV confined matter. Each throat has its own inverse $\ads$ curvature denoted by $k_1$ and $k_2$.}
\label{modelpics}
\end{figure}

As a toy example and proof of principle, we will assume that the axion arises as a Nambu-Goldstone mode associated with a higher dimensional complex scalar field which spontaneously breaks a $\mathrm{U(1)}_{PQ}$ symmetry,
\be
\phi \approx \frac{\tilde f_{PQ}}{\sqrt{2}} e^{ia/\tilde f_{PQ}}.
\ee
We will further assume that the vev remains constant over the entire $\ads$ slice. This scenario can arise through a simple multi-throat model where the standard model exists in its own throat (SM throat) while the complex scalar exists in a separate throat (PQ throat) as illustrated in figure \ref{modelpics}. In the PQ throat we have,
\be
S_\Phi = M_5\int d^5 x \sqrt{|g|} g^{MN} \partial_M \Phi^* \partial_N \Phi
\ee
where upon decomposing $\Phi = \eta(y) \exp(ia(y))$ we apply the boundary conditions
\bea
\left.\eta\right|_{IR} = v&& \hspace{4mm} \left.\eta\right|_{UV} = v \nonumber \\
\left.\partial_ a\right|_{IR} = 0&& \hspace{4mm} \left.\partial_z a\right|_{UV} = 0
\eea
which break the global $U(1)_{PQ}$ symmetry and yield a constant solution $\langle \eta \rangle = v$ for the vev. The resulting axion, $a(y)$, couples to the standard model via UV brane-localized interactions.\footnote{In \cite{WA}, the UV brane localized interactions are mediated by exotic coloured fermions, $\bar Q, Q$, with action $S_Q$$=$$\int d^4 x \sqrt{|g_\mathrm{ind}|_{UV}} (\Phi \bar Q_L Q_R + \Phi^* \bar Q_R Q_L)$. Under a $U(1)$ chiral rotation of the $Q$ fields, the axion coupling becomes transfered to a gluon topological term of QCD on the UV brane, namely
$S_{\mathrm{int}}$$=$$\int d^4 x \spc (32\pi^2 f_{PQ})^{-1} a F_{\mu\nu}\tilde F^{\mu\nu}$ which corresponds to the generic UV localized axion coupling we use in this article.} 

While generically we also expect a KK tower associated with the radial mode, we do not allow any radial mode-standard model interactions. Furthermore, the radial mode interactions appear through derivative couplings suppressed by powers of $\tilde f_{PQ}$. As we will show later in this section, these couplings are further suppressed by a volume factor. As a result, for the remainder of this paper we will ignore the effects of the radial mode. 

We stress that the above toy model only represents a proof of principle. We will take a model independent approach in that we will not consider any particular implementation of the PQ symmetry breaking mechanism within the $\ads$ set-up other than insisting on the constancy of the vev across the $\ads$ slice. In this paper, we only wish to explore the phenomenology of a bulk axion in a warped geometry. We will assume that the standard model remains confined to the UV brane and that the axion propagates in the bulk (as in  figure \ref{modelpics} (b)). Since we assume that the axion potential only arises from QCD instanton effects, we will omit bulk or boundary mass terms for the axion. Defining the line element within the $\ads$ space as,
\be
ds^2 = e^{-2ky} \eta_{\mu\nu} dx^\mu dx^\nu + dy^2 \equiv g_{MN} dx^M dx^N
\ee
($\eta_{\mu\nu} = \mathrm{diag}(-1,1,1,1)$) our action reads,
\be
S_{eff} = \int d^4x dy \spc e^{-4ky} \left[\mathcal{L}_{\mathrm{SM}}\delta(y) + \frac{1}{2} g^{M N} \partial_M a\partial_N a + \frac{g^2\xi}{32\pi^2}\frac{a}{\tilde f_{PQ}} F_{\mu\nu}\tilde F^{\mu\nu} \delta(y)\right].
\ee
The axion field, $a(x,y)$, carries mass dimension $[3/2]$ implying that $[\tilde f_{PQ}] =3/2$. The five dimensional Newton's constant has been absorbed into the field definitions. Application of Neumann boundary conditions in the absence of bulk or boundary axion mass terms leads to the KK decomposition,
\bea
a(x,y) &=& \frac{1}{\sqrt{2\pi R}} \sum_{n=0} a_n(x) \phi_n(y) \\
&=& \frac{1}{\sqrt{2\pi R}} a_0(x) \phi_0 + \frac{1}{\sqrt{2\pi R}} \sum_{n=1} a_n \phi_0(y)
\eea
where $\phi_0$ is a constant. Decomposing the action leads to,
\be
S = \int d^4x dy \spc e^{-4ky}\left[e^{2ky}\frac{1}{2}\eta^{\mu\nu}\partial_\mu a \partial_\nu a + \frac{1}{2}(\partial_5 a)^2 + \frac{g^2\xi}{32\pi^2}\frac{a}{\tilde f_{pq}} F_{\mu\nu}\tilde F^{\mu\nu} \delta(y)\right].
\ee
The kinetic terms,
\be
\partial_\mu a \partial^\mu a = \frac{1}{2\pi R} \left(\partial_\mu \sum_{n=0} a_n\phi_n\right)\left(\partial^\mu \sum_{n=0} a_n\phi_n\right),
\ee
contain no cross terms since the wave function profiles satisfy the orthogonality condition
\be
\frac{1}{\pi R} \int_0^{\pi R} dy \spc e^{-4ky} e^{2ky} \phi_m \phi_n = \delta_{mn}
\ee
for all $n,m$ -- including the zero mode. We may write the wave function profiles as,
\be
\phi_n = \frac{e^{2ky}}{N_n}\left[J_2 \left(\frac{m_n}{k} e^{ky}\right) + b^{(n)}_2 Y_2\left(\frac{m_n}{k} e^{ky}\right)\right]
\ee
for $n >0$ with the normalization factor,
\be
N_n^2  = \frac{1}{\pi R} \int_0^{\pi R} dy \spc\left[J_2\left(\frac{m_n}{k}e^{ky}\right) + b^{(n)}_2 Y_2\left(\frac{m_n}{k}e^{ky}\right)\right]^2.
\ee
The coefficients $b_2^{(n)}$ are determined by,
\bea
b_2^{(n)} &=& - \frac{2 J_2(m_n/k) + (m_n/k) J^\prime_2(m_n/k)}{2 Y_2(m_n/k) + (m_n/k) Y^\prime_2(m_n/k)} \\
b_2^{(n)}(m_n) &=& b_2^{(n)}(m_n e^{\pi kR})
\eea
and we now can re-express the $4$-D action as
\be
S = \int d^4 x\spc\left[\frac{1}{2}\sum_{n=0}\partial_\mu a_n\partial^\mu a_n - \frac{1}{2} \sum_{n=1} m_n^2 (a_n)^2 + \frac{g^2\xi}{32\pi^2}\frac{1}{\sqrt{2\pi R}} \frac{1}{\tilde f_{pq}} F_{\mu\nu}\tilde F^{\mu\nu}\sum_{n=0}a_n \phi_n(0)\right].
\ee
At this point, we will now make the definition,
\be
\hat f_{pq} = \sqrt{2\pi R} \tilde f_{pq}
\label{hat}
\ee
leading to the parameter $\hat f_{pq}$ which has canonical mass dimension 1. The parameter $\tilde f_{PQ}$ implicitly depends on the 5-dimensional Newton's constant. Thus, a volume factor relates the two parameters in eq.(\ref{hat}). This allows us to set $\hat f_{PQ}$ to a hierarchically high scale.

Like the flat extra dimensional counterpart \cite{DDG}, this construction solves the strong CP problem. Using the one-instanton dilute gas approximation, we have
\be
\langle F^{\mu\nu}\tilde F_{\mu\nu}\rangle = -\Lam^4 \sin\left(\frac{\xi}{\hat f_{pq}} \sum_{n=0} a_n \phi_n(0) + \bar \Theta\right)
\ee
which gives rise to the axion potential,
\be
V = \frac{1}{2}\sum_{n=1} m_n^2 (a_n)^2 + \frac{g^2}{32\pi^2} \Lam^4\left[1 - \cos\left(\frac{\xi}{\hat f_{pq}} \sum_{n=0} a_n \phi_n(0) + \bar \Theta\right)\right].
\ee
The CP conserving minimum appears from,
\be
0=\frac{\partial V}{\partial a_n} = m_n^2 a_n + \phi_n \frac{\xi}{\hat f_{pq}} \frac{g^2}{32\pi^2} \Lam^4\sin\left(\frac{\xi}{\hat f_{pq}} \sum_{n=0} a_n \phi_n(0) + \bar \Theta\right)
\label{axi_pot}
\ee
leading to the condition:
\bea
\langle a_0\rangle &=& \frac{\hat f_{pq}}{\xi} \frac{l\pi-\bar\Theta}{\phi^{(0)}(0)} \hspace{5mm} l \in 2 \mathbb{Z} \nonumber \\
\langle a_n \rangle &=& 0  \hspace{6mm} n >0.
\eea
We see that only the zero mode serves as the true axion. The higher dimensional axion mass matrix obtained from eq.(\ref{axi_pot}) reads,
\bea
M_{mn}^2 &=& \frac{\partial^2 V}{\partial a_m \partial a_n} \\
&=& m_n^2 \delta_{mn} + \frac{g^2 \xi^2}{32\pi^2} \frac{\Lam^4}{\hat f^2_{pq}}\phi_m(0)\phi_n(0)\left.\cos\left(\frac{\xi}{\hat f_{pq}} \sum_{n=0} a_n \phi_n(0) +\bar\Theta\right)\right|_{\langle a\rangle}
\eea
and therefore we obtain,
\be
M_{mn}^2 = m_{PQ}^2\left(\phi_m(0) \phi_n(0) + y_n^2 \delta_{mn}\right)
\label{mass_matrix}
\ee
with
\bea
m_{PQ}^2 &=& \frac{g^2\xi^2}{32\pi^2}\frac{\Lam^4}{\hat f_{pq}^2} \\
y_0^2 &=& 0 \\
y_n^2 &=& \frac{m_n^2}{m_{PQ}^2} \hspace{3mm} n>0.
\eea
We can re-express $\phi^{(m)}(0)$ using appropriate Bessel function identities,
\be
b^{(n)}_2 = - \frac{J_1(m_n/k)}{Y_1(m_n/k)}
\ee
yielding
\be
\phi_n(0) = \frac{1}{N_n}\left[J_{2}\left(\frac{m_n}{k}\right) - \frac{J_1(m_n/k)}{Y_1(m_n/k)} Y_2\left(\frac{m_n}{k}\right)\right].
\ee

\subsection{Axion KK mode mixing}

The important difference between the usual 4-dimensional axion and its higher dimensional implementation centers on the effects of KK mode mixing given in eq.(\ref{mass_matrix}). It has been shown \cite{DDG} that flat compactifications can lead to a divorcing of the axion zero mode mass from the PQ scale, leading to the approximate relation
\be
m_0 \approx \mathrm{min}\left( \frac{1}{2} R^{-1}, m_{PQ}\right)
\ee
for the lightest axionic state. The mixing of the axion KK modes, via the mass matrix eq.(\ref{mass_matrix}), leads to important differences between the physics of the usual four-dimensional axion and its higher-dimensional counterpart. In the standard four-dimensional axion scenario, the PQ scale not only sets the mass of the axion but also controls the strength of the axion coupling to the standard model. Observational constraints imply a PQ scale between $10^{9}$ GeV and $10^{12}$ GeV corresponding to an axion mass window of $10^{-5}$ eV$\lesssim m_{PQ} \lesssim 10^{-2}$ eV. In the extra dimensional situation, where the axion appears as a bulk field, these standard arguments no longer apply: the would-be axion $a_0$ mixes with the KK tower via the mass matrix eq.(\ref{mass_matrix}). This mixing implies that $m_{PQ}$ no longer sets the lightest axion state. Using the property that the smallest eigenvalue of a real symmetric matrix is bounded above by the eigenvalues of any diagonal block, one can show (from the upper-left $2\times2$ diagonal block of eq.(\ref{mass_matrix})) that the mass of the lightest axionic state is bounded above by both $m_{PQ}$ and $m_1$. Hence, $m_a \leq \mathrm{min} (m_{PQ},m_1)$, and we see that for $m_{PQ} \gtrsim m_1$, the lightest axion mass decouples from $m_{PQ}$ (and therefore $\hat f_{PQ}$) - the lightest KK mass can set the scale. This result parallels the flat extra-dimensional case (see eq.(47) of \cite{DDG}) where the inverse radius of the extra dimension can also set the lightest axion mass. 

Once the compactification radius and $f_{PQ}$ are chosen within the flat compactification scenario, the form of eq.(\ref{mass_matrix}) becomes fixed and leads to a definite KK mixing pattern. Warped compactifications lead to more freedom. In the large $kR$ limit ($kR \gg 1$), we have 
\be
N_n \approx \frac{e^{\pi kR/2}}{\sqrt{\pi^2 R m_n}},
\ee
and, using asymptotic expansions for all the Bessel functions involved, we obtain
\be
\phi_n(0) \approx -\pi e^{-\pi kR/2}\sqrt{R m_n} \left[1 - \frac{1}{8} \left(\frac{m_n}{k}\right)^2\right].
\ee
The mass matrix eq.(\ref{mass_matrix}) becomes
{\tiny\be
M^2 = m_{PQ}^2\left( \begin{array}{ccccc}
2\pi kR & -\sqrt{2\pi^3 kR^2 m_1}e^{-\pi kR/2} & -\sqrt{2\pi^3 kR^2 m_2}e^{-\pi kR/2}& -\sqrt{2\pi^3 kR^2 m_3}e^{-\pi kR/2}&... \\
-\sqrt{2\pi^3 kR^2 m_1}e^{-\pi kR/2} & e^{-\pi kR}\pi^2(m_1R) + \frac{m_1^2}{m_{PQ}^2} & e^{-\pi kR}\pi^2R\sqrt{m_1m_2}& e^{-\pi kR}\pi^2R\sqrt{m_1m_3} &...  \\
-\sqrt{2\pi^3 kR^2 m_2}e^{-\pi kR/2} & e^{-\pi kR}\pi^2R\sqrt{m_2m_1}  & e^{-\pi kR}\pi^2(m_2R) + \frac{m_2^2}{m_{PQ}^2}  & e^{-\pi kR}\pi^2R\sqrt{m_2m_3}  & ... \\
-\sqrt{2\pi^3 kR^2 m_3}e^{-\pi kR/2} & e^{-\pi kR}\pi^2R\sqrt{m_3m_1}  & e^{-\pi kR}\pi^2R\sqrt{m_3m_2} & e^{-\pi kR}\pi^2(m_3R) + \frac{m_3^2}{m_{PQ}^2} &... \\
... & ... & ... &... &...
\end{array} \right).
\label{upper}\ee}
This pattern substantially differs from the flat compactification scenario. The main new feature concerns the mass of the zero mode. Like the flat 5-dimensional case, the axion does not receive a mass term until $m_{PQ}$ turns on. By contrast, in the highly warped case, the zero mode ``axion" contributes a mass eigenvalue of
\be
m_0 \approx m_{PQ}\sqrt{2\pi kR}.
\ee
In particular, this mass eigenvalue can be tuned relative to the first several KK states. Furthermore, the amount of mixing between the zero mode and the other KK states becomes tunable through the warping, $kR$. We note that if we choose $kR \ll 1$ at fixed $R$, we recover the flat higher dimensional results. 

We should also emphasize the nature of the mixing matrix at large KK number. In this case,
\be
b_2^{(n)} = -\frac{J_1(m_n/k)}{Y_1(m_n/k)} \approx  -\cot(m_n/k -3\pi/4) \Rightarrow m_n = \frac{n\pi k}{e^{\pi kR}-1}
\ee
and we find,
\be
N^2_n \approx \frac{1}{\pi^2 R m_n} \csc^2\left(\frac{m_n}{k} - \frac{3\pi}{4}\right) e^{\pi kR}
\ee
which leads to
\be
\phi_n \approx \sqrt{2} \left(\sqrt{\pi Rk} e^{-\pi kR/2}\right) \hspace{4mm} n \gg 1.
\label{assy}
\ee
In this limit, eq.(\ref{mass_matrix}) now reads,

{\tiny\be
M^2 = m_{PQ}^2 (\pi Rk) e^{-\pi kR}\left( \begin{array}{ccccc}
... &... &... &... &... \\
... & 2 + \frac{\pi n^2(k/R) e^{-\pi kR}}{m_{PQ}^2} & 2 & 2 & 2  \\
... & 2 & 2 + \frac{\pi (n+1)^2(k/R) e^{-\pi kR}}{m_{PQ}^2} & 2 & 2 \\
... & 2 & 2 & 2 + \frac{\pi (n+2)^2(k/R) e^{-\pi kR}}{m_{PQ}^2} & 2 \\
... & 2 & 2 & 2 & 2 + \frac{\pi (n+3)^2(k/R) e^{-\pi kR}}{m_{PQ}^2}
\end{array} \right).
\label{lower}\ee}
Apart from the exponential suppression factors, the mixing matrix approaches the flat case form. We stress that the $\phi_n$ coefficients approach a constant as $n\rightarrow \infty$. This feature will play an important role in the following sections.

\section{Decoherence, Invisibility, Energy Loss, and Detection}

Perhaps the greatest potential threat to extra dimensional axion scenarios concerns the loss of invisibility. While each individual KK mode couples with $1/\hat f_{PQ}$ (as in \cite{DDG}), the ``axion'' is produced in a linear superposition of KK modes. The effective coupling to this superposition determines the axion's invisibility. Therefore, the effective interaction Lagrangian reads,
\be
\mathcal{L}_{int} = \frac{g^2\xi}{32\pi^2} \sqrt{N} \frac{a^\prime}{\hat f_{PQ}} F^{\mu\nu a}\tilde F_{\mu\nu a}
\label{int_eff}
\ee
where $N$ is a normalization factor defined by,
\be
N = \sum_{n=0}^{n_{max}} (\phi_n)^2
\label{normalize}
\ee
and
\be
a^\prime = \frac{1}{\sqrt{N}}\sum_{n=0}^{n_{max}} \phi_n a_n
\ee
defines the ``axion''. We stress that only the zero mode state, $a^0$, serves as the true axion in this set up as only $a^0$ inherits the shift symmetry from the five dimensional theory. Since the entire linear superposition of the KK tower given by $a^\prime$ enters the effective Lagrangian, the KK sum must be truncated -- presumably at the cutoff of the 5-D effective field theory. However, processes which produce on-shell axions should not include KK states above the characteristic production scale. 

If we consider axion processes at characteristic energy $E$, the effective field theory which results from integrating out axion mass eigenstates that exceed $E$ reads, $\mathcal{L}_{eff} = \Sigma_{n=0}^{n_{\mathrm{max}}}(\partial a_n)^2 +\Sigma_{m,n=0}^{n_{\mathrm{max}}} M_{mn} a_m a_n + \frac{g^2 \xi}{32 \pi^2 \hat{f_{PQ}}} \Sigma_{n=0}^{n_{\mathrm{max}}} \phi_n a_n$. Ignoring the the effects of mixing from eq.(\ref{mass_matrix}), we can approximate $n_{\mathrm{max}}$ through $m_{n_{\mathrm{max}}} \simeq E$. We see that the gauge fields couple to the normalized linear superposition $a^\prime = (N(E))^{-1/2} \Sigma_{n=0}^{n_{\mathrm{max}}} \phi_n a_n $ with coupling strength  $\sqrt{N(E)}\hat{f}_{PQ}^{-1}$. The normalization factor $N(E) = \Sigma_{n=0}^{n_{\mathrm{max}}} \phi_n^2$ enhances the effective coupling. For clarity we have indicated the implicit energy dependence of $N$ arising from the condition $m_{n_{\mathrm{max}}} \simeq E$. Thus, the effective coupling strength of the linear superposition $a^\prime$ depends on the characteristic energy scale as does the composition of $a^\prime$, and the determination of axion invisibility becomes much more subtle in an extra-dimensional scenario. Furthermore, the state $a^\prime$ produced in interactions with visible matter is manifestly not a mass eigenstate, and so neither its mass nor its lifetime are well-defined.

Since $N$ depends on KK number, and since the kinematics of a given process fix the number of modes included in $a^\prime$, the effective coupling in eq.(\ref{int_eff}) grows with energy. In general, eq.(\ref{mass_matrix}) creates a misalignment between mass and interaction eigenstates such that
\be
\hat a_{l} = \sum_{n=0}^{n_{max}} U_{l n}a_n,
\label{mis}
\ee
where $\hat a$ denotes the mass eigenstate and $U$ diagonalizes the axion mass matrix, eq.(\ref{mass_matrix}). From the 4-dimensional perspective, at a given energy scale, $a^\prime$ appears as a field in the interaction basis. As the state $a^\prime$ propagates, the individual KK modes interfere, creating an axion oscillation phenomenon. Again, we stress that $a^\prime$ consists of a superposition of mass eigenstates and does not have a well-defined lifetime or mass. We will examine two important physical consequences: energy loss and direct detection.

The strongest bounds on the 4-dimensional Peccei-Quinn scale arise from stellar cooling constraints. Weakly coupled axions provide new channels for stellar energy loss and the total luminosity in axions must not upset stellar evolution. The current lower bound on the Peccei-Quinn scale in the usual 4-dimensional axion scenario from HB stellar cooling reads (see \cite{Raffelt} and references therein),
\be
f_{PQ} \gtrsim 2 \times 10^{9}\spc \mathrm{GeV}.
\label{lowerbound}
\ee 
This bound applies provided that the axion mass does not wildly exceed the internal temperature of the star and the coupling remains sufficiently weak \cite{Raffelt}. In the higher dimensional analogue, the effective coupling contains energy dependence from the inclusion of KK modes. Thus, eq.(\ref{lowerbound}) becomes modified, leading to
\be
\frac{\hat{f}_{PQ}}{\sqrt{N(E)}} \gtrsim 2\times 10^{9}\spc \mathrm{GeV}
\ee
where $N(E)$ denotes the normalization factor of eq.(\ref{normalize}) which includes modes up to $n_{max}$. For simplicity, we will assume that the mode superposition within $a^\prime$ contains states up to a cutoff defined through the internal stellar temperature via $m_{n_{max}} = T$. We will also make the further approximation that the usual 4-dimensional kinematical phase space constraints apply for $a^\prime$ production.\footnote{Strictly speaking, we expect that thermal corrections to the phase space will play a role in the production of the more massive states in the superposition, as shown in \cite{Pilaftsis}. Since we crudely truncate the tower at $T$, for the purposes of this paper, we ignore these effects.}

In the warped case, we can provide an estimate on the energy dependence of $N(E)$. If we approximate the mass spectrum by the KK masses (again, ignoring the effect of eq.(\ref{mass_matrix}) on the mass eigenvalues), at large KK number we have
\be
E \approx m_{n_{max}} \approx \pi k e^{-\pi kR}
\ee
which leads to
\be
n_{max} \approx \frac{E}{\pi k} e^{\pi kR}.
\ee
From eq.(\ref{normalize}), and using the asymptotic approximation for all $\phi_n$, eq.(\ref{assy}), we can obtain an approximate explicit formulation of $N(E)$ in terms of the kinematic cutoff,
\bea
N(E) &\approx& (2\pi kR)\left(1 + e^{-\pi kR}\sum_{n=1}^{n_{max}} (1)\right) \nonumber \\
&\approx& 2 E R.
\eea
Thus, at a fixed energy, the effective Peccei-Quinn scale for $a^\prime$ becomes
\be
\left.\hat f^{eff}_{PQ}\right|_{E} \approx \frac{\hat f_{PQ}}{\sqrt{2R E}}
\ee
where we assume $RE \gg 1$. In order to satisfy HB stellar evolution constraints (assuming core temperatures of $\sim 8\, \mathrm{keV}$ \cite{Raffelt}), we require
\be
\left.\hat f^{eff}_{PQ}\right|_{E = 8\spc\mathrm{keV}} \gtrsim 2\times 10^{9}\spc\mathrm{GeV}
\ee 
implying,
\be
\frac{\hat f_{PQ}}{\sqrt{(2R)\cdot 8\spc\mathrm{keV}}} \gtrsim 2\times 10^{9} \spc\mathrm{GeV}.
\ee
Note the above result obtained for a warped compactification gives comparable bounds for flat compactification \cite{DDG,Pilaftsis}: 
\be
n_{max} = ER
\ee
and thus,
\be
N_E  = 1 + \sum_{n=1}^{n_{max}} 2 \approx 2 E R.
\ee
Both the flat and warped compactifications give the same normalization factor at fixed energy.

While the enhanced coupling must satisfy the astrophysical energy loss constraints, the effects of the superposition within the $a^\prime$ field lead to phenomenological consequences in direct detection searches. Again, the relevant quantity is not the individual KK modes nor the suppressed coupling $\hat f_{PQ}$, but the superposition $a^\prime$ with coupling $\hat f^{eff}_{PQ}$. Even though the extra-dimensional scenario leads to an enhanced coupling, the KK modes interfere during propagation. Only the linear superposition $a^\prime$ couples to standard model fields. It has been shown \cite{DDG} in flat compactifications that destructive interference is crucial in re-establishing invisibility as it reduces the expected flux measured from distant sources. 

In order to determine what fraction of the flux produced by a distant source can be detected, we must calculate the survival probability of the $a^\prime$ field itself. The amplitude for individual KK axion transitions is given by,
\be
A_{k\rightarrow l}(t) = \sum_{i} U_{i l} U^*_{i k} \exp\left(-i\frac{m_i^2}{2p} t\right)
\label{trans}
\ee
where $m_i$ denotes the mass eigenvalue arising from eq.(\ref{mass_matrix}), and $U$ labels the unitary matrix that diagonalizes the axion mass matrix. We have ignored the possibility of axion decay throughout. The probability that $a^\prime$ remains preserved during propagation reads,
\be
P_{a^\prime\rightarrow a^\prime}(t) = \frac{1}{N^2}\left|\sum_{k,l} \phi_l \phi_k A_{k\rightarrow l}(t)\right|^2.
\label{survive}
\ee
Therefore, the expected measurable flux arriving from a distant source appears as,
\be
\Phi = \langle P_{a^\prime\rightarrow a^\prime} \rangle \Phi_0
\ee
where $\Phi_0$ denotes the flux as calculated using the enhanced coupling $\hat f^{eff}_{PQ}$ in the absence of decoherence. Since $\langle P_{a^\prime\rightarrow a^\prime} \rangle \leq 1$ we see that interference serves to reduce the the total measurable flux.

As an estimate of the decoherence time we consider the time it takes for the largest mass eigenstate in the linear superposition, $a^\prime$, to step completely out-of-phase with the zero mode. Applying this out-of-phase condition to eq.(\ref{trans}) yields,
\be
0= \cos\left(\frac{m_{n}^2 -m_0^2}{2p}t_0\right)
\label{heavy_mode}
\ee
and we find that,
\be
\tau_0^* \approx \frac{m_{PQ}^2}{(n\pi k)^2} e^{2\pi kR}.
\label{min_1}
\ee
The parameter $\tau_0^*$, defined through $\tau_0^* = t_0(m^2_{PQ}/2p)$ (where $p$ denotes the usual 3-momentum), gives a dimensionless time element.
In the flat case limit, eq.(\ref{heavy_mode}) yields,
\be
\tau_0^* \approx \left(\frac{m_{PQ} R}{n}\right)^2
\label{min_2}
\ee
in agreement with \cite{DDG}.

Note that $\tau_0^*$ denotes a ``decoherence time" defined by $P_{a^\prime\rightarrow a^\prime}(\tau_0^*) \sim \langle P_{a^\prime\rightarrow a^\prime} \rangle$ -- i.e. the time taken for $P_{a^\prime\rightarrow a^\prime}(t)$ to reach its time independent average value. The scaling argument proceeds subtly. We know that the axion remains in the $a^\prime$ state as $m_{PQ} \rightarrow 0$ since the mass matrix of eq.(\ref{mass_matrix}) becomes diagonal in this limit and, as a result, no mode mixing can occur (i.e. the transition amplitudes become diagonal, $A_{k\rightarrow l} = \delta_{kl}$). This leads to the complete lack of decoherence and reproduces the usual 4-dimensional axion scenario. As $m_{PQ}\rightarrow 0$ we find that $\tau_0^* \rightarrow 0$, implying a vanishing decoherence time. This result remains consistent since $P_{a^\prime\rightarrow a^\prime}(\tau) \sim \langle P_{a^\prime\rightarrow a^\prime} \rangle \rightarrow 1$ in this limit. Thus, even though $\tau_0^* \rightarrow 0$ as $m_{PQ} \rightarrow 0$ we also have $\langle P_{a^\prime\rightarrow a^\prime} \rangle \rightarrow 1$ and hence decoherence does not occur.\footnote{We stress that our definition of decoherence differs from \cite{DDG} where the authors define decoherence by fixing $P_{a^\prime\rightarrow a^\prime}(\tau_0)= 0.90$ for all values of $m_{PQ}$ and $R$. Their definition requires a non-trivial mode number renormalization in which the energy dependence of $f^{eff}_{PQ}$, of the $a'$ composition -- and therefore the energy dependence of $P_{a^\prime\rightarrow a^\prime}$ -- is not manifest.}

The largest KK mass in both the flat and warped cases govern eq.(\ref{min_1}) and eq.(\ref{min_2}). We may re-write the decoherence time as
\be
\tau_0^* \approx \left(\frac{m_{PQ}}{m^{(n)}_{KK}}\right)^2,
\ee
which holds for both cases. Approximating the energy of the linear superposition through the mass of the heaviest mode we find,
\be
\tau_0^* \approx \left(\frac{m_{PQ}}{E}\right)^2
\label{indy}
\ee
which is independent of the details of the compactification. Thus, provided that many modes appear in the superposition, large warped and flat compactifications lead to the same decoherence time at fixed energy.

Furthermore, we can obtain a semi-analytic expression for the resulting time independent average value $\langle P_{a^\prime\rightarrow a^\prime} \rangle$. In the flat compactification at fixed $m_{PQ}$ and fixed $R$, it has been shown \cite{DDG} that $\langle P_{a^\prime\rightarrow a^\prime} \rangle \sim 1/n$. This behaviour follows from expanding eq.(\ref{trans}) and eq.(\ref{survive}). Applying these results to the flat compactification scenario gives
\be
P_{a^\prime\rightarrow a^\prime} \sim \frac{1}{n}\left[a \sum_{i} + 2 b \sum_{i<j} \cos \left(\frac{[m_i^2 - m_j^2]t}{2p}\right)\right].
\label{expa}
\ee
The $\mathcal{O}(1)$ coefficients $a$ and $b$ result from from the unitary matrix that diagonalizes the flat extra-dimensional axion mass matrix. At $t=0$ eq.({\ref{expa}}) gives the expected result 
\be
P_{a^\prime\rightarrow a^\prime} \sim \left(\frac{1}{n^2}\right) n^2 \approx 1.
\ee
At later times, the cosine terms no longer add coherently and thus eq.(\ref{expa}) reduces to
\be
P_{a^\prime\rightarrow a^\prime} \sim \left(\frac{1}{n^2}\right) n \sim \frac{1}{n}.
\label{flat_n}
\ee

The warped case proceeds in a similar fashion. By observing that eq.(\ref{lower}) has the same structure as the resulting flat compactification matrix (apart from an overall multiplicative factor), we see that eq.(\ref{expa}) applies. In particular, at large mode number, the essential difference between the warped and flat compactifications centers on the density of KK states. At a fixed energy, the warped compactification contains $\sim \pi kR \exp(\pi kR)$ more modes than the flat case. Eq.(\ref{expa}) then tells us the scaling in the warped case appears as,
\be 
P_{a^\prime\rightarrow a^\prime} \sim \frac{e^{\pi kR}}{\pi kR} \left(\frac{1}{n}\right)
\label{warped_n}
\ee
indicating an exponentially larger value for the expectation of $P_{a^\prime\rightarrow a^\prime}$ at fixed $n$ as compared to the flat compactification result. However, if we compare the results eq.(\ref{flat_n}) and eq.(\ref{warped_n}) not at fixed $n$, but at fixed energy we find that in both cases
\be
P_{a^\prime\rightarrow a^\prime} \sim \frac{1}{ER},
\label{warped_scale}
\ee
since the warped case yields the approximate relation $E \sim n \pi kR \exp(\pi kR)$. Remarkably, at fixed energy, both the warped and flat extra dimensional axion scenarios lead to a similar value for $\langle P_{a^\prime\rightarrow a^\prime}\rangle$. Thus, while it appears that at fixed mode number warped compactifications permit decoherence times that are substantially longer than those occurring in the flat case limit, physical processes cutoff the mode number in both cases to yield a result that is largely independent of the compactification.

As examples, we show the effect of warping on the decoherence time relative to flat compactifications in figures \ref{decohere_1} and \ref{decohere_2}. In both cases, we have taken $R = 1/(10^{-2}\spc \mathrm{eV})~\approx~10\spc~\mu\mathrm{m}$, $\hat f_{PQ} = 10^{11}\spc\mathrm{GeV}$. 

In figure~\ref{decohere_1}, where a fixed number of 30 KK modes are used, it appears as though the reduced mixing in the warped compactification scenario (see eq.({\ref{mass_matrix})) reduces the axion's ability to decohere. Figure~\ref{decohere_1}(a) displays the effect with a mild warp factor of $kR=0.5$. In this case, we find that the flat compactification large time limit yields $\langle P_{a^\prime\rightarrow a^\prime} \rangle \approx 3.3 \times 10^{-2}$ while the warped compactification gives a similar value of $\langle P_{a^\prime\rightarrow a^\prime}\rangle\approx 3.9\times 10^{-2}$. In figure~\ref{decohere_1}(b) we use a larger warp factor of $kR = 1$. The flat compactification reproduces $\langle P_{a^\prime\rightarrow a^\prime} \rangle \approx 3.3\times 10^{-2}$ in the large time limit. On the other hand, the warped compactification large time limit yields $\langle P_{a^\prime\rightarrow a^\prime} \rangle \approx 0.23$, marking a significant departure from the flat compactification result \emph{at fixed mode number}.

\begin{figure}[ht!]
\newlength{\picwidthc}
\setlength{\picwidthc}{3in}
 \begin{center}
\subfigure[][]{\resizebox{\picwidthc}{!}{\includegraphics{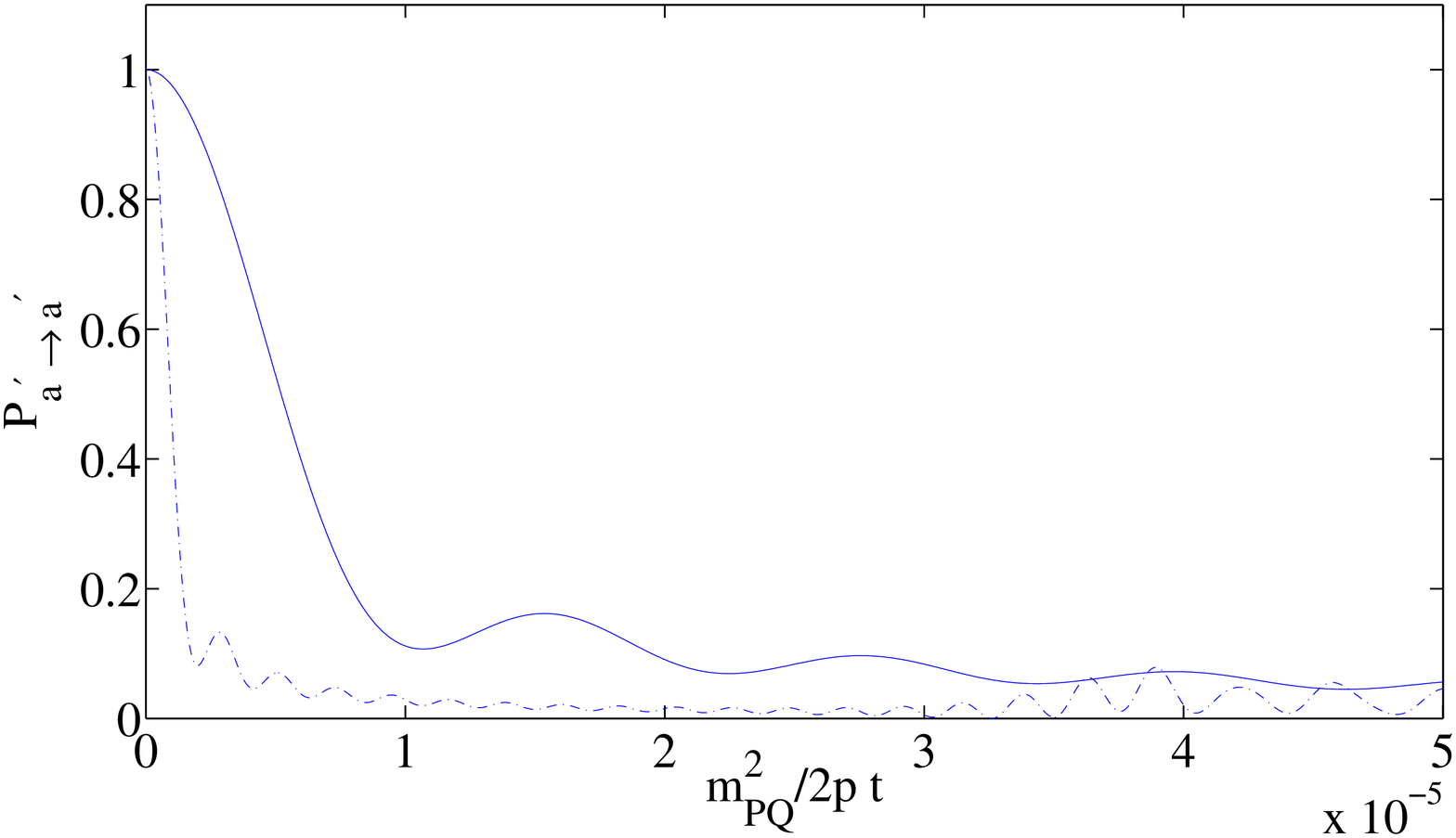}}}
\subfigure[][]{\resizebox{\picwidthc}{!}{\includegraphics{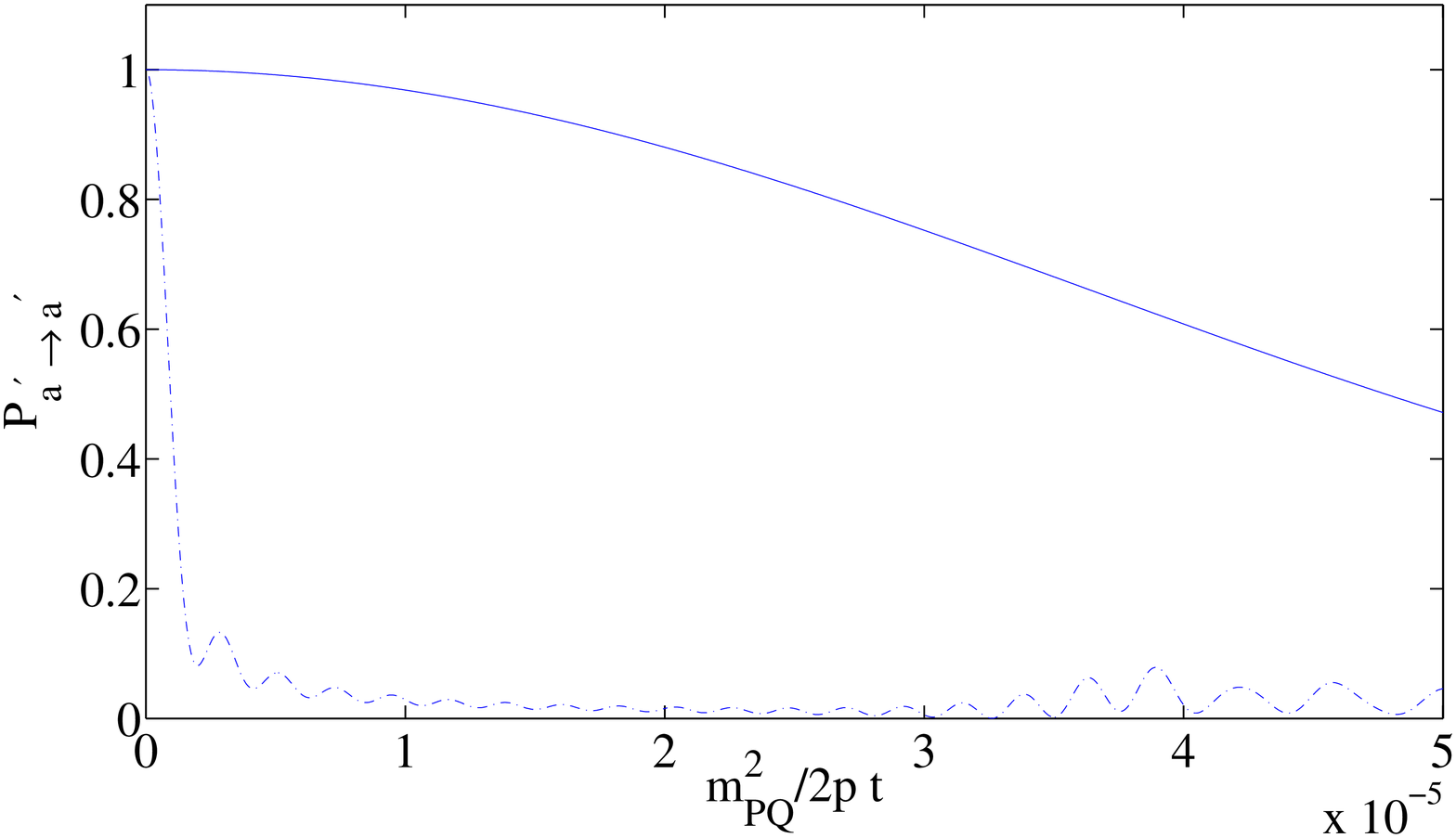}}}
%\subfigure[][]{\resizebox{\picwidthc}{!}{\includegraphics{kR_6_warped_wide.ps}}}
 \end{center}
\caption{The axion survival probability with $kR = 0.5$ and $kR = 1$ with $30$ KK modes included. Plots a) and b) displays the flat compactification (dashed-dotted) along with the warped compactification (solid line). In all plots $R = 10^{2} \spc\mathrm{eV}^{-1}$, and $\hat f_{PQ} = 10^{11}\spc\mathrm{GeV}$.}
\label{decohere_1}
\end{figure}

However, in \emph{physical} processes, the maximum mode number within the linear superposition is fixed by kinematics. While the axion mass matrix in the warped case appears with suppressed mixing relative to the flat case, the KK spectrum also appears more dense. The increase in mode number in the warped case makes up for the suppressed mixing leading to eq.(\ref{warped_scale}) -- i.e. the same result as derived in the flat compactification scenario. Figure \ref{decohere_2} displays the results at fixed energy, again for $kR=0.5$ and $kR=1$. In this case we have taken the mode number cutoff determined through $E = 0.1$ eV in both cases. We see that $\tau_0^*$ and the final value for $\langle P_{a^\prime\rightarrow a^\prime} \rangle$ appear approximately the same.

\begin{figure}[ht!]
\newlength{\picwidthd}
\setlength{\picwidthd}{3in}
 \begin{center}
\subfigure[][]{\resizebox{\picwidthd}{!}{\includegraphics{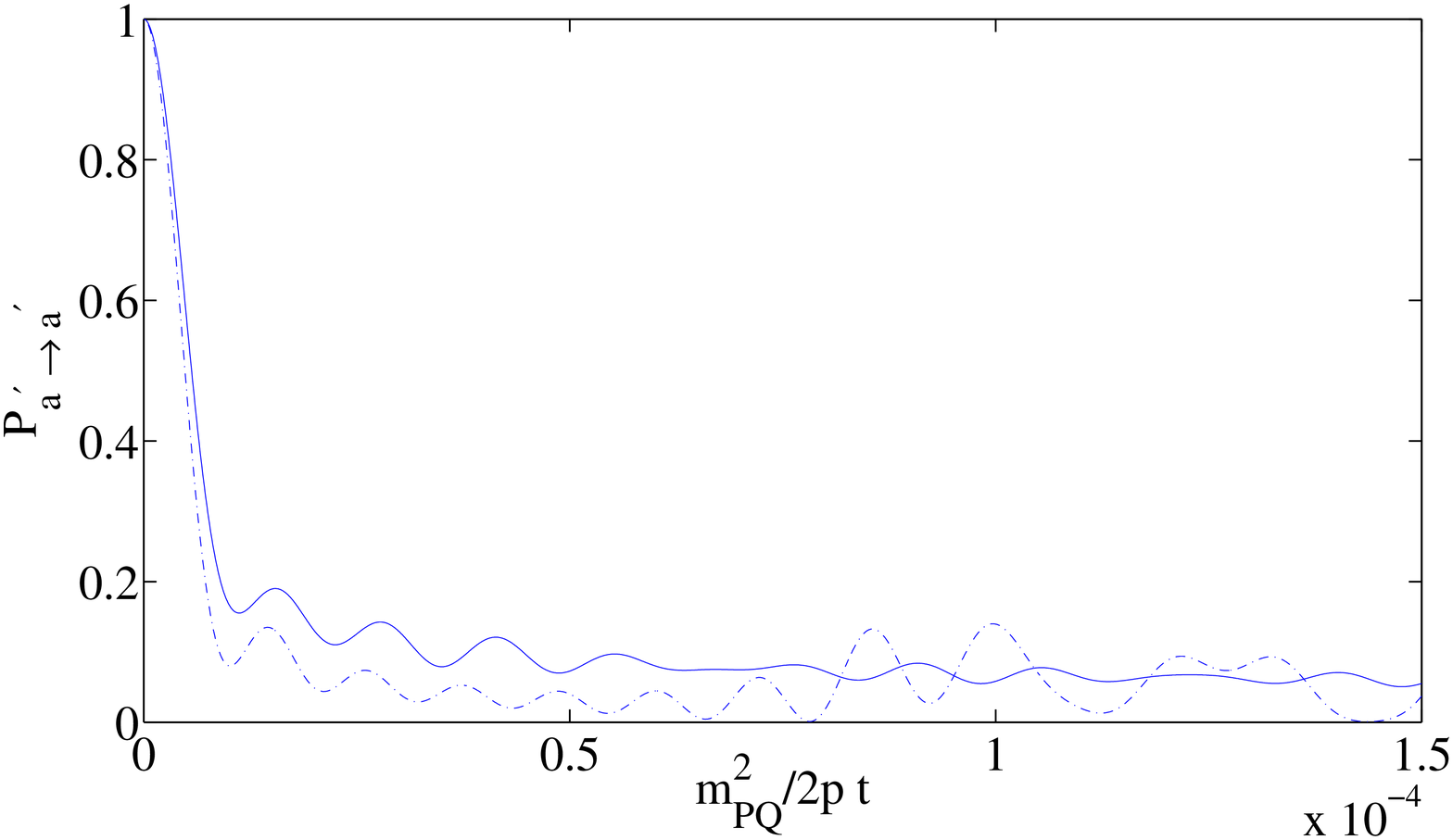}}}
\subfigure[][]{\resizebox{\picwidthd}{!}{\includegraphics{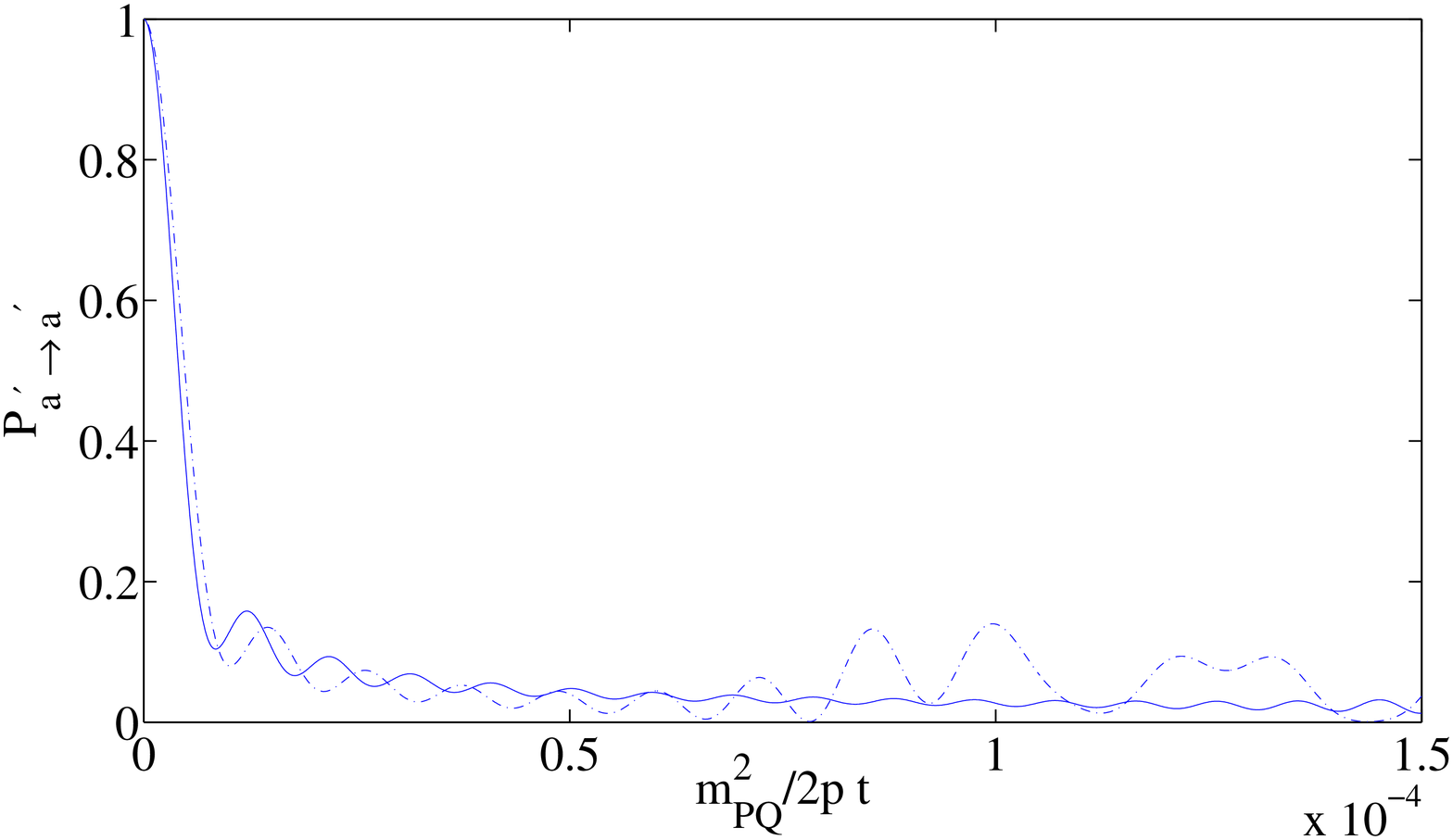}}}
%\subfigure[][]{\resizebox{\picwidthc}{!}{\includegraphics{kR_6_warped_wide.ps}}}
 \end{center}
\caption{The axion survival probability with $kR = 0.5$ and $kR = 1$ at a fixed energy of $0.1$ eV. Plots a) and b) displays the flat compactification (dashed-dotted) along with the warped compactification (solid line). In all plots $R = 10^{2} \spc\mathrm{eV}^{-1}$, and $\hat f_{PQ} = 10^{11}\spc\mathrm{GeV}$. Note that $\tau_0^* \approx (m_{pq}/E)^2$ and $\langle P_{a^\prime\rightarrow a^\prime} \rangle \approx 1/ER$ in both cases.}
\label{decohere_2}
\end{figure}

\section{Comment on Cosmological Relic Axions}

Cosmological relic axions provide strong constraints on the invisibility of the axion in the usual 4-dimensional PQ mechanism. During the universe's early thermal history, the universe passed through the QCD phase transition, at time $t_{QCD}$, at which point QCD instanton effects established an axion potential where none existed previously. By naturalness, we expect that the axion would find itself displaced from the minimum of its potential by an $\mathcal{O}(1)$ fraction of $\hat f_{PQ}$. At this point, the axion would begin to oscillate about the minimum via
\be
\frac{d^2 a}{dt^2} + 3 H(t) \frac{d a}{dt} + m_a^2 a = 0. \hspace{4mm} t> t_{QCD}
\label{oscte_1}
\ee
with the initial condition $a(t_{QCD}) \sim \hat f_{PQ}$. These relic oscillations, while damped through Hubble expansion, continue to store energy and must not exceed the present day critical density. Cosmological considerations provide a constraint from above on $\hat f_{PQ}$ in the usual 4-dimensional PQ mechanism, namely $\hat f_{PQ} \lesssim 10^{12}\spc\mathrm{GeV}$. 

\begin{figure}[ht!]
\newlength{\picwidthz}
\setlength{\picwidthz}{3in}
 \begin{center}
\subfigure[][]{\resizebox{\picwidthz}{!}{\includegraphics{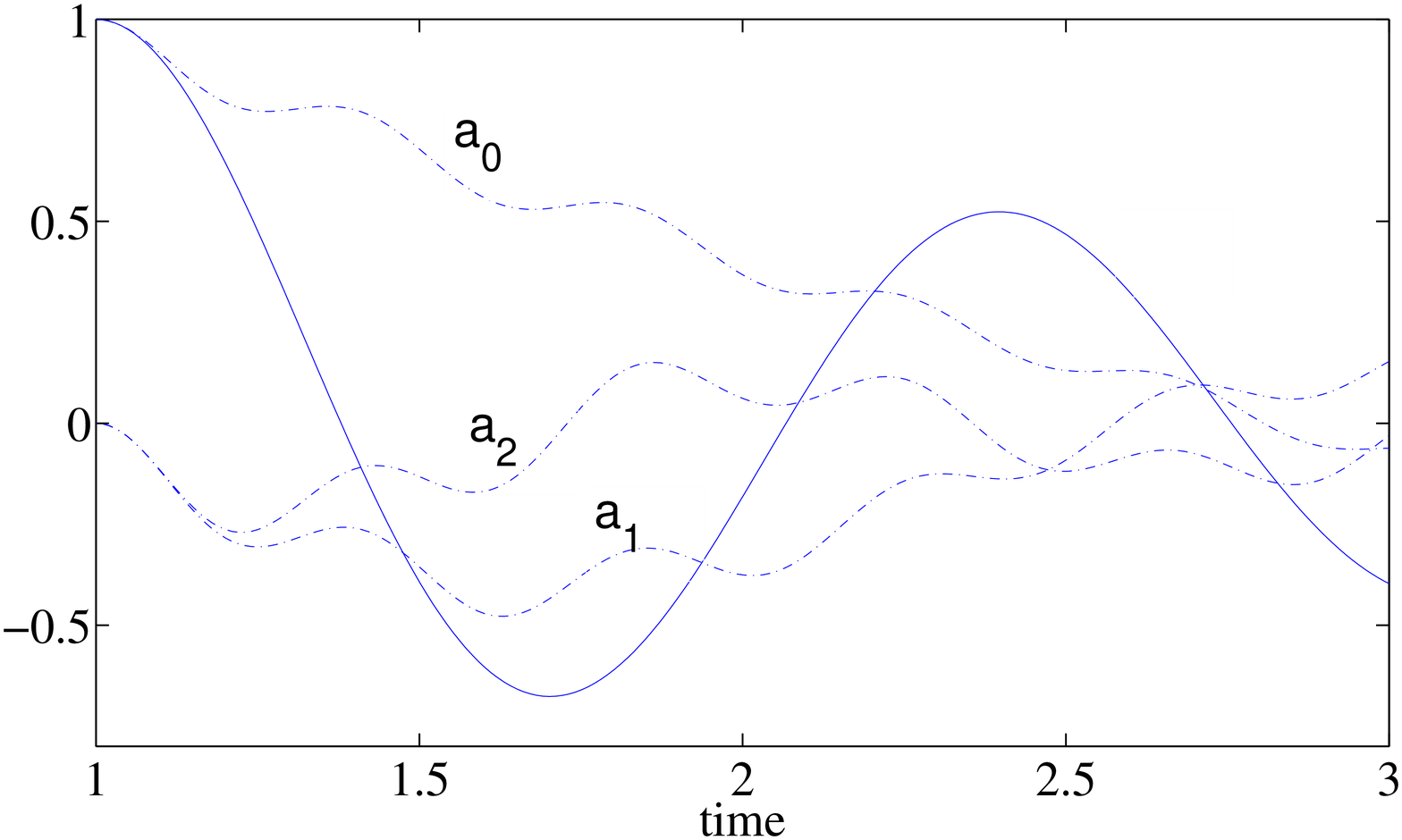}}}
\subfigure[][]{\resizebox{\picwidthz}{!}{\includegraphics{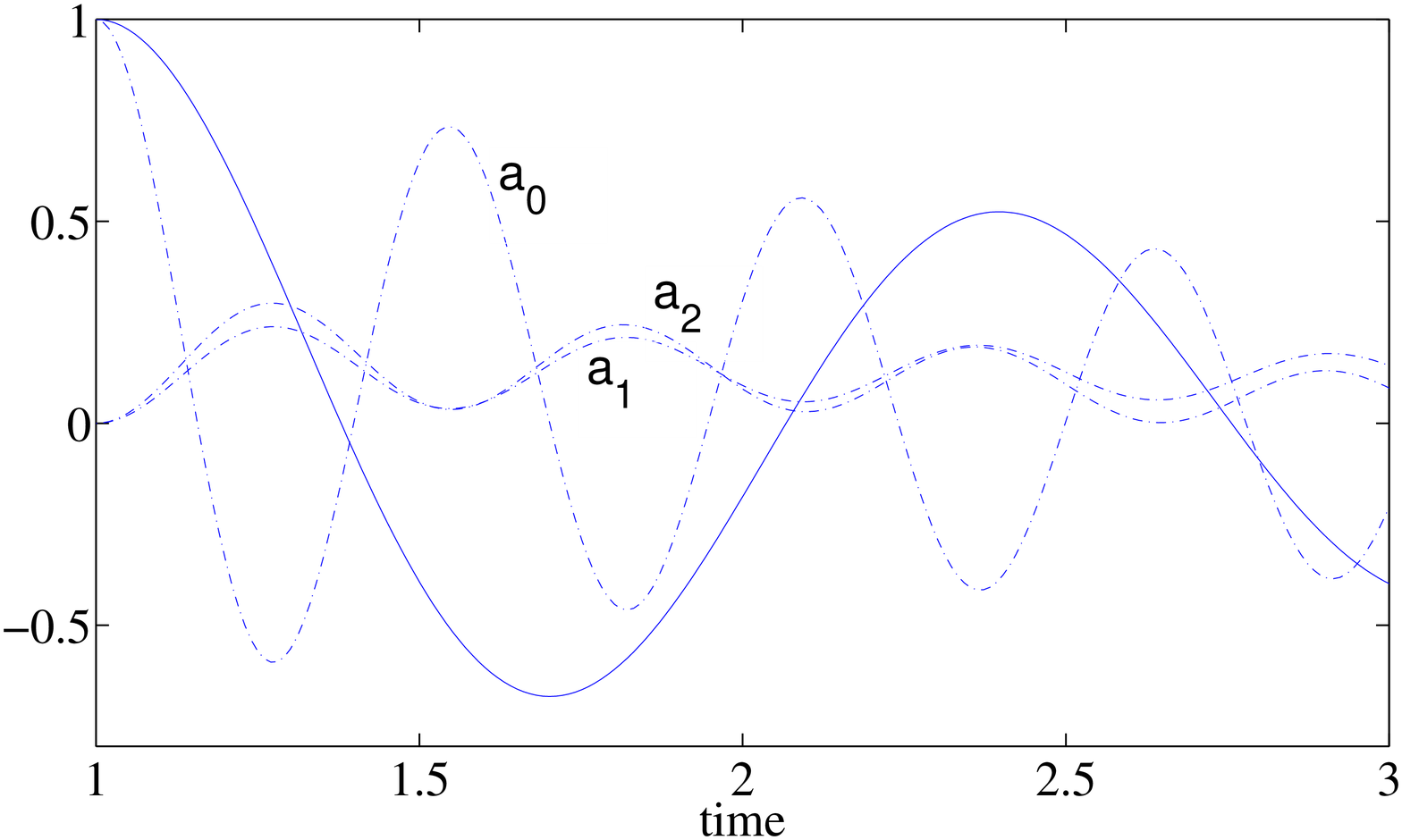}}}
 \end{center}
\caption{Higher dimensional induced KK oscillation plot as a function of time (dimensionless) with $t_{QCD} = 1$; $m_{pq} = 4\times 10^{-4}\spc\mathrm{eV}$, $R = 1/(2\times10^{-4}\spc\mathrm{eV})$.  Plot a) shows the first three KK mode oscillation with flat compactification (dashed-dotted) lines. The solid line denotes the usual 4-dimensional axion scenario. Plot b) displays the effect of warping with $kR =2$ among the first three modes. Again, the solid line denotes the usual 4-dimensional axion scenario.}
\label{osc_1}
\end{figure}

As displayed in figure~\ref{osc_1}, the extra dimensional situation proceeds differently. As each KK mode has a potential prior to the QCD phase transition and following \cite{DDG}, we can approximate each KK mode as sitting at the minimum of its potential at $t_{QCD}$. By contrast, the zero mode (which represents the true axion) does not receive a potential until the instanton effects turn on at $t_{QCD}$. Thus, we expect at $t_{QCD}$ that the zero mode will have a natural displacement from its minimum, i.e. $a_0(t_{QCD}) \sim \hat f_{PQ}$, $a_l(t_{QCD}) =0$ for $l>0$, and $da_l/dt = 0$ for all $l$. In the extra dimensional case, eq.(\ref{oscte_1}) reads,
\be
\frac{d^2 a_l}{dt^2} + 3 H(t) \frac{d a_l}{dt} + M_{lk}^2 a_k = 0 \hspace{4mm} t> t_{QCD}
\label{oscte_2}
\ee
where $M_{lk}^2$ denotes the non-diagonal axion mass matrix. Once the zero mode begins to oscillate, the non-diagonal mass matrix in eq.(\ref{oscte_2}) induces sympathetic KK mode oscillations that potentially have the ability to alter the cosmological constraints obtained in the usual 4-dimensional case. It has been shown in \cite{DDG} that for certain parameter choices the oscillating KK modes can dissipate the oscillation energy more quickly relative to the 4-dimensional case, allowing $\hat f_{PQ}$ to become as large as $\sim 10^{16}\spc\mathrm{GeV}$. However, once constraints from low energy phenomenology and gravity are applied, it can be shown \cite{DDG} that the extra-dimensional situation dissipates the oscillation energy at the same rate as the usual 4-dimensional scenario. Remarkably, cosmological constraints on the flat the extra-dimensional axion scenario lead to the same viability as the usual 4-dimensional axion.

Following \cite{DDG}, the results of the flat extra-dimensional case can be readily extended to warped compactifications. We can transform to the mass eigenbasis such that eq.(\ref{oscte_2}) appears as a set of uncoupled ordinary differential equations,
\be
\frac{d^2 \tilde a_l}{d t^2} + \frac{3}{2t} \frac{d\tilde a_l}{dt} + m_l^2 \tilde a_l =0,
\label{uncoupled}
\ee
where we have assumed a radiation dominated epoch, $H(t) = 3/2t$, and we have defined $\tilde a_l = \hat a_l/\hat{f}_{PQ}$. For each uncoupled differential equation in eq.(\ref{uncoupled}), we can use the dimensionless time element $\tau = m_l t$ leading to,
\be
\frac{d^2 \tilde a_l}{d \tau^2} + \frac{3}{2\tau} \frac{d\tilde a_l}{d\tau} + \tilde a_l =0,
\label{uncoupled2}
\ee
The initial conditions now read,
\be
\tilde a_l(\tau_0) = U_{0l}, \hspace{4mm} \left.\frac{d\tilde a_l}{d\tau}\right|_{\tau=\tau_0} =0
\label{initial_con}
\ee
where $U$ diagonalizes the axion mass matrix, eq.(\ref{mass_matrix}). Using the initial conditions in eq.(\ref{initial_con}), the general solution to eq.(\ref{uncoupled2}) reads,
\be
\tilde a_l(\tau) = -\frac{\pi}{\sqrt{2}} U_{0l} \tau_0^{5/4} \tau^{-1/4} j(\tau_0;\tau)
\label{solution}
\ee
where 
\be
j(\tau_0;\tau) = J_{-5/4}(\tau_0) J_{1/4}(\tau) + J_{5/4}(\tau_0) J_{-1/4}(\tau).
\ee
This solution matches the results given in \cite{DDG} up to the unitary matrix element, $U_{0l}$. Defining the dimensionless mass eigenvalue $\tilde m_l = m_l/m_{PQ}$ the total energy density can be written as,
\be
\tilde \rho(\tau) = \sum_l \frac{\tilde m_l^2}{2}\left(\tilde a_l^2 + \left(\frac{d \tilde a_l}{d\tau}\right)^2\right)
\ee
with $\tilde \rho \equiv \rho/(m^2_{PQ} \hat f_{PQ}^2)$. More succinctly, by defining $\tilde t = m_{PQ} t$, the total energy density becomes,
\be
\tilde \rho(\tilde t) = \frac{\pi^2}{4} \tilde t_0^{5/2} \tilde t^{-1/2} \sum_l U_{0l}^2 \tilde m_l^4\left(j(\tilde m_l \tilde t_0;\tilde m_l \tilde t)^2 + j^\prime(\tilde m_l \tilde t_0;\tilde m_l \tilde t)^2\right).
\label{total_full}
\ee
In the large time limit $\tilde m_l \tilde t \gg 1$ eq.(\ref{total_full}) becomes,
\be
\tilde \rho (\tilde t) = \frac{\pi}{2} X(\tilde t_0) \tilde t_0 ^{5/2} \tilde t^{-3/2}
\label{as}
\ee
where the time independent coefficient $X(\tilde t_0)$ reads
\be
X(\tilde t_0) = \sum U_{0l}^2 \tilde m_l^3\left([J_{5/4}(\tilde m_l \tilde t_0)]^2 + [J_{-5/4}(\tilde m_l \tilde t_0)]^2 + \sqrt{2}J_{5/4}(\tilde m_l \tilde t_0) J_{-5/4}(\tilde m_l \tilde t_0)  \right).
\label{X}
\ee
Again, this result appears the same as in the flat compactification \cite{DDG} up to the unitary matrix element. Since the energy density for the four dimensional case in the large time limit has the same form as eq.(\ref{as}) except with
\be
X_{4D}(t_0) = [J_{5/4}(\tilde t_0)]^2 + [J_{-5/4}(\tilde t_0)]^2 + \sqrt{2} J_{5/4}(\tilde t_0) J_{-5/4}(\tilde t_0)
\ee
the ratio of the energy densities can be expressed as  \cite{DDG},
\be
r_\rho \equiv \frac{\rho(t)}{\rho_{4D}(t)} = \frac{X(\tilde t_0)}{X_{4D}(\tilde t_0)}.
\label{energy_ratio}
\ee

As in the flat case, $r_\rho$ can significantly deviate from unity leading to a weakening of the upper bound on $\hat f_{PQ}$. This deviation occurs only in the limit where $\tilde t_0 \tilde m_l < \mathcal{O}(1)$ (i.e. the Bessel functions in eq.(\ref{X}) do not approximate their asymptotic limit) for a significant number of modes which also includes the mass eigenvalue contributed by the mode $a_0$. Nevertheless, $r_\rho$ saturates at unity which implies that extra-dimensional axions -- warped or flat -- remain viable cosmologically. In figure \ref{ratios} we have displayed three cases, $\tilde t_0 = 0.1, 1, 10$, with $m_{PQ}R =1$.  We find that as the warping becomes increased, the deviation of $r_\rho$ from unity increases as well, provided that the mass eigenvalue contributed by $a_0$ satisfies $\tilde t_0 m <\mathcal{O}(1)$. While the plots in figure \ref{ratios} demonstrate the potential ability of warping to further increase the viability of the extra dimensional axion scenario, we caution that the large departure of $r_\rho$ from unity crucially hinges on certain parameter choices. In the flat case, the condition required for $r_\rho <1$ reads $t_{QCD}/R \lesssim \mathcal{O}(1)$ -- a condition requiring $R\sim 10^{10}\spc \mbox{eV}^{-1}$! This result wildly contradicts constraints from gravity. The warped case proceeds more subtly. If the zero-mode in eq.(\ref{mass_matrix}) contributes a typical mass eigenvalue of $\sim k\pi\exp(-\pi kR)$ to the KK spectrum then, provided that $t_{QCD} k \pi \exp(-\pi k R) \lesssim \mathcal{O}(1)$, a departure of $r_\rho$ from unity can be expected. However, in the large warping limit ($kR \gtrsim 1$), the zero-mode clearly contributes $\sim m_{PQ} \sqrt{2\pi kR}$ (see eq.(\ref{mass_matrix})) to the KK spectrum thereby requiring $t_{QCD} (m_{PQ} \sqrt{2\pi kR}) \lesssim \mathcal{O}(1)$ in order for us to expect $r_\rho < 1$. Using $t_{QCD} \approx 10^{-5}\spc\mathrm{sec} \approx 10^{10} \spc \mathrm{eV}^{-1}$, and $kR \sim 1 $ with $R \sim 10\spc\mu\mathrm{m}$, we find no significant deviation of $r_\rho$ from unity unless $m_{PQ} \lesssim 10^{-10}\spc\mathrm{eV}$ (which generally does not include an enhancement factor larger than $\mathcal{O}(1)$). Thus, over the parameter range of interest, we can expect that warped extra dimensional axions dissipate the relic oscillation energy no more quickly that the usual four dimensional scenario. This result matches the flat extra dimensional case. We stress the remarkable central point: the extra dimensional axion scenario -- whether warped or flat -- continues to remain cosmologically viable, dissipating the relic oscillation energy at at rate no less than the usual four dimensional axion scenario. 

\begin{figure}[ht!]
\newlength{\picwidthe}
\setlength{\picwidthe}{4.5in}
 \begin{center}
{\resizebox{\picwidthe}{!}{\includegraphics{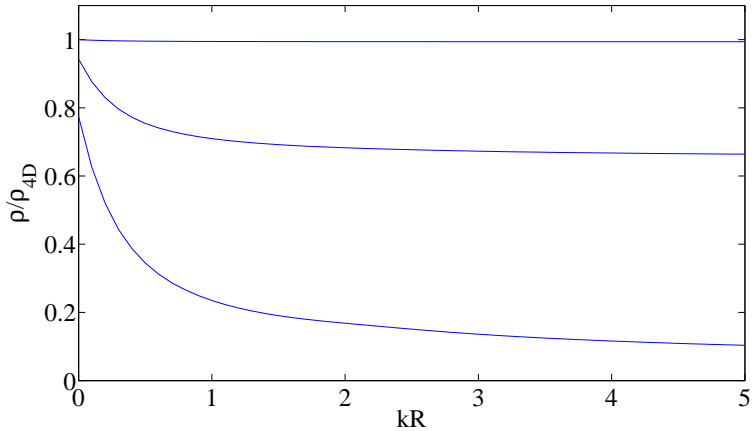}}}
 \end{center}
\caption{Ratio of the energy densities for $\tilde t_0 =0.1$,$1$,$10$ as a function of the warp factor kR with $m_{PQ} = 1/R$.  The $k=0$ limit yields the flat case results in \cite{DDG}.}
\label{ratios}
\end{figure}

\section{Conclusions}

In this paper, we have compared the basic phenomenology of axions embedded in a flat extra dimension with axions embedded in a warped extra dimension. Given the recent interest in multi-throat scenarios \cite{multithroats}, large warped extra dimensions may provide new model building avenues. Furthermore, current and planned experimental axion searches will probe new parameter ranges in axion physics \cite{AxExperiments}. 

In a flat extra dimension, axion oscillations cause decoherence which leads to invisibility. We have found that warping the extra dimension controls the axion oscillations such that the decoherence length becomes tunable in KK number. While this result appears to give greater freedom in establishing the level of axion invisibility, kinematics govern the number of modes that must be included in the axion linear superposition. As the warped case demands that we include more modes relative to the flat case at a given parameter choice, we surprisingly find that the overall effect leaves the invisibility unchanged relative to flat compactifications.
 
Furthermore, in both warped and flat extra dimensions, kinematics also govern the effective coupling, $\hat f^{eff}_{PQ}$. While each mode couples with safe $\sim \phi_n/\hat f_{PQ}$, the on-shell state couples to the standard model with the much larger $\hat f^{eff}_{PQ} \sim f_{PQ}/\sqrt{N(E)}$, where $N(E)$ is a function of the production energy. Thus, the production kinematics play a central role in determining axion's coupling to the standard model -- for both warped and flat compactifications. Constraints derived from energy loss mechanisms must take this effect into account -- it is not the coupling of an individual mode the establishes the bound but rather the coupling of the superposition. Within extra-dimensional scenarios, the energy dependent effective coupling, $\hat f^{eff}_{PQ}$, controls astrophysical axion production, and therefore controls the source's axion luminosity. On the other hand, decoherence, arising from KK axion oscillations, sets the expected measurable flux. In the warped geometry, the AdS curvature, $k$, plays a crucial role in both effects and yet for processes that involve large mode numbers, the expected phenomenology remains essentially unaltered from flat compactifications. It would be interesting to perform a detailed analysis involving constraints from astrophysical sources on axions in the warped extra dimension scenario we consider.

We have also found that warped compactifications remain cosmologically viable. As in the flat case, we have found that the sympathetic KK mode oscillations induced by the displaced zero-mode dissipate the axion oscillation energy at least the same rate as the usual four-dimensional axion over most of the parameter range. If we include a large warp factor $(kR \gtrsim 1)$ and take $m_{PQ}t_{QCD} \lesssim \mathcal{O}(1)$, we find the possibility that the KK modes dissipate the relic oscillation energy more quickly than the four-dimensional axion scenario. While this result in principle allows a larger value of $\hat f_{PQ}$ relative to the usual four-dimensional bound $(\sim 10^{12} \spc\mathrm{GeV})$, the resulting factor by which the dissipation becomes increased is generally insufficient to allow a significant increase of $\hat f_{PQ}$.

\section{Acknowledgments}
We would like to thank John March-Russell and Ben Gripaios for their contribution at an earlier stage of this work. We also wish to thank Gianmassimo Tasinato and Martin Schvellinger for useful discussions. The work of TF is supported by ``Evangelisches Studienwerk Villigst e.V.'' and PPARC (Grant No PPA/S/S/2002/03540A). DM wishes to acknowledge the support of the Natural Sciences and Engineering Research Council of Canada. This work was also partially supported by the EC 6th Framework Programme MRTN-CT-2004-503369.


\begin{thebibliography}{99}
\bibitem{Eidelman:2004wy}
  S.~Eidelman {\it et al.}  [Particle Data Group],
  %``Review of particle physics,''
  Phys.\ Lett.\ B {\bf 592} (2004) 1.
  %%CITATION = PHLTA,B592,1;%%
\bibitem{Weinberg:1975ui}
  S.~Weinberg,
  %``The U(1) Problem,''
  Phys.\ Rev.\ D {\bf 11}, 3583 (1975).
  %%CITATION = PHRVA,D11,3583;%%
\bibitem{'tHooft:1976up}
  G.~'t Hooft,
  %``Symmetry breaking through Bell-Jackiw anomalies,''
  Phys.\ Rev.\ Lett.\  {\bf 37} (1976) 8.
  %%CITATION = PRLTA,37,8;%%
\bibitem{PDG}
  W.~M.~Yao {\it et al.}  [Particle Data Group],
  ``Review of particle physics,''
  J.\ Phys.\ G {\bf 33}, 1 (2006).
  %%CITATION = JPHGB,G33,1;%%
\bibitem{Peccei:1988ci}
  R.~D.~Peccei,
  %``The Strong CP Problem,''
  Adv.\ Ser.\ Direct.\ High Energy Phys.\  {\bf 3}, 503 (1989).
\bibitem{Peccei:2006as}
  R.~D.~Peccei,
  %``The strong CP problem and axions,''
  arXiv:hep-ph/0607268.
  %%CITATION = HEP-PH 0607268;%%
\bibitem{Kim:1986ax}
  J.~E.~Kim,
  %``LIGHT PSEUDOSCALARS, PARTICLE PHYSICS AND COSMOLOGY,''
  Phys.\ Rept.\  {\bf 150}, 1 (1987).
  %%CITATION = PRPLC,150,1;%%\ Rev.\ Lett.\  {\bf 37}, 8 (1976).
  %%CITATION = PRLTA,37,8;%%
\bibitem{Peccei:1977hh}
  R.~D.~Peccei and H.~R.~Quinn,
  %``CP Conservation In The Presence Of Instantons,''
  Phys.\ Rev.\ Lett.\  {\bf 38}, 1440 (1977).
  %%CITATION = PRLTA,38,1440;%%
\bibitem{ww}
  S. Weinberg, Phys. Rev. Lett. {\bf 40}, 223 (1978);\spc
  F. Wilczek, Phys. Rev. Lett. {\bf 40}, 279 (1978).
\bibitem{Bardeen_and_Tye}
  W.~A.~Bardeen and S.~H.~Tye,
  %``Current Algebra Applied To Properties Of The Light Higgs Boson,''
  Phys.\ Lett.\ B {\bf 74} (1978) 229;\spc
  %%CITATION = PHLTA,B74,229;%%
  J.~Kandaswamy, P.~Salomonson and J.~Schechter,
  %``Mass Of The Axion,''
  Phys.\ Rev.\ D {\bf 17} (1978) 3051.
  %%CITATION = PHRVA,D17,3051;%%
\bibitem{Raffelt}
  G.~G.~Raffelt,
  %``Astrophysical methods to constrain axions and other novel particle
  %phenomena,''
  Phys.\ Rept.\  {\bf 198} (1990) 1.
  %%CITATION = PRPLC,198,1;%%
\bibitem{KSVZ}
  J. E. Kim, Phys. Rev. Lett. {\bf 43}, 103 (1979); 
  M. A. Shifman, V. I. Vainshtein and V. I. Zakharov,
  Nucl. Phys. {\bf B166}, 4933 (1980).
\bibitem{DFSZ}
  M. Dine, W. Fischler and M. Srednicki, Phys.
  Lett. {\bf B104}, 199 (1981);\spc
  A.~R.~Zhitnitsky,
  %``On Possible Suppression Of The Axion Hadron Interactions. (In Russian),''
  Sov.\ J.\ Nucl.\ Phys.\  {\bf 31}, 260 (1980)
  [Yad.\ Fiz.\  {\bf 31}, 497 (1980)].
  %%CITATION = SJNCA,31,260;%%
\bibitem{DDGprecursors}
  N.~Arkani-Hamed, S.~Dimopoulos and G.~R.~Dvali,
  %``The hierarchy problem and new dimensions at a millimeter,''
  Phys.\ Lett.\ B {\bf 429} (1998) 263
  [arXiv:hep-ph/9803315];\spc
  %%CITATION = HEP-PH 9803315;%%
  % N.~Arkani-Hamed, S.~Dimopoulos and G.~R.~Dvali,
  %``Phenomenology, astrophysics and cosmology of theories with  sub-millimeter
  %dimensions and TeV scale quantum gravity,''
  Phys.\ Rev.\ D {\bf 59} (1999) 086004
  [arXiv:hep-ph/9807344];\spc
  %%CITATION = HEP-PH 9807344;%%
  I.~Antoniadis {\it et al.}
  %``New dimensions at a millimeter to a Fermi and superstrings at a TeV,''
  Phys.\ Lett.\ B {\bf 436} (1998) 257
  [arXiv:hep-ph/9804398];\spc
  %%CITATION = HEP-PH 9804398;%%
  S.~Chang, S.~Tazawa and M.~Yamaguchi,
  %``Axion model in extra dimensions with TeV scale gravity,''
  Phys.\ Rev.\ D {\bf 61} (2000) 084005
  [arXiv:hep-ph/9908515].
  %%CITATION = HEP-PH 9908515;%%
\bibitem{DDG}
  K.~R.~Dienes, E.~Dudas and T.~Gherghetta,
  %``Invisible axions and large-radius compactifications,''
  Phys.\ Rev.\ D {\bf 62}, 105023 (2000)
  [arXiv:hep-ph/9912455].
  %%CITATION = HEP-PH 9912455;%%
\bibitem{RS}
  L.~Randall and R.~Sundrum,
  %``An alternative to compactification,''
  Phys.\ Rev.\ Lett.\  {\bf 83} (1999) 4690
  [arXiv:hep-th/9906064];\spc
  %%CITATION = HEP-TH 9906064;
  %``A large mass hierarchy from a small extra dimension,''
  Phys.\ Rev.\ Lett.\  {\bf 83} (1999) 3370
  [arXiv:hep-ph/9905221].
  %%CITATION = HEP-PH 9905221;%%
\bibitem{bottomup}
  K.~W.~Choi,
  %``A QCD axion from higher dimensional gauge field,''
  Phys.\ Rev.\ Lett.\  {\bf 92} (2004) 101602
  [arXiv:hep-ph/0308024];\spc
  %%CITATION = HEP-PH 0308024;%%
  H.~Collins and R.~Holman,
  %``The invisible axion in a Randall-Sundrum universe,''
  Phys.\ Rev.\ D {\bf 67}, 105004 (2003)
  [arXiv:hep-ph/0210110].
  %%CITATION = HEP-PH 0210110;%%    
\bibitem{stringaxions}
  See eg., P.~Svrcek and E.~Witten,
  %``Axions in string theory,''
  JHEP {\bf 0606}, 051 (2006)
  [arXiv:hep-th/0605206], and references therein.
%%CITATION = HEP-TH 0605206;%%
\bibitem{conlon}
  J.~P.~Conlon,
  %``The QCD axion and moduli stabilisation,''
  JHEP {\bf 0605}, 078 (2006)
  [arXiv:hep-th/0602233];\spc
  %%CITATION = HEP-TH 0602233;%%
  I.~W.~Kim and J.~E.~Kim,
  %``Modification of decay constants of superstring axions: Effects of flux
  %compactification and axion mixing,''
  Phys.\ Lett.\ B {\bf 639} (2006) 342
  [arXiv:hep-th/0605256].
  %%CITATION = HEP-TH 0605256;%%
\bibitem{multithroats}
  S.~Dimopoulos, S.~Kachru, N.~Kaloper, A.~E.~Lawrence and E.~Silverstein,
  %``Small numbers from tunneling between brane throats,''
  Phys.\ Rev.\ D {\bf 64}, 121702 (2001)
  [arXiv:hep-th/0104239];\spc
  %%CITATION = HEP-TH 0104239;%% 
  % S.~Dimopoulos, S.~Kachru, N.~Kaloper, A.~E.~Lawrence and E.~Silverstein,
  %``Generating small numbers by tunneling in multi-throat  compactifications,''
  Int.\ J.\ Mod.\ Phys.\ A {\bf 19}, 2657 (2004)
  [arXiv:hep-th/0106128];\spc
  %%CITATION = HEP-TH 0412040;%%
  G.~Cacciapaglia, C.~Csaki, C.~Grojean and J.~Terning,
  %``Field theory on multi-throat backgrounds,''
  arXiv:hep-ph/0604218;\spc
  %%CITATION = HEP-PH 0604218;%%
  %%CITATION = HEP-TH 0106128;%%
  A.~Hebecker and J.~March-Russell,
  %``The ubiquitous throat,''
  arXiv:hep-th/0607120.
  %%CITATION = HEP-TH 0607120;%%
\bibitem{WA}
  T.~Flacke, B.~Gripaios, J.~March-Russell and D.~Maybury,
  %``Warped axions,''
  arXiv:hep-ph/0611278.
  %%CITATION = HEP-PH 0611278;%%
\bibitem{Pilaftsis}
  L.~Di Lella, A.~Pilaftsis, G.~Raffelt and K.~Zioutas,
  %``Search for solar Kaluza-Klein axions in theories of low-scale quantum
  %gravity,''
  Phys.\ Rev.\ D {\bf 62} (2000) 125011
  [arXiv:hep-ph/0006327].
  %%CITATION = HEP-PH 0006327;%%
\bibitem{AxExperiments}
  K.~Zioutas {\it et al.}  [CAST Collaboration],
  %``First results from the CERN axion solar telescope (CAST),''
  Phys.\ Rev.\ Lett.\  {\bf 94} (2005) 121301
  [arXiv:hep-ex/0411033];\\
  %%CITATION = HEP-EX 0411033;%%
  S.~J.~Asztalos {\it et al.} [ADMX Collaboration],
  %``An improved RF cavity search for halo axions,''
  Phys.\ Rev.\ D {\bf 69} (2004) 011101
  [arXiv:astro-ph/0310042];\\
  %%CITATION = ASTRO-PH 0310042;%%
  E.~Zavattini {\it et al.}  [PVLAS Collaboration],
  %``Experimental observation of optical rotation generated in vacuum by a
  %magnetic field,''
  Phys.\ Rev.\ Lett.\  {\bf 96} (2006) 110406
  [arXiv:hep-ex/0507107];\\
  %%CITATION = HEP-EX 0507107;%%
  G.~Cantatore [PVLAS Collabortion],
  ``Probing the quantum vacuum with polarized light: a low energy photon-photon collider at PVLAS,"
  2nd ILIAS-CERN-CAST Axion Academic Training 2006,
  http://cast.mppmu.mpg.de/\\
  A.~V.~Afanasev, O.~K.~Baker and K.~W.~McFarlane [LIPSS Collaboration],
  % ``Production and detection of very light spin-zero bosons at optical
  %frequencies,''
  %arXiv:
  arXiv:hep-ph/0605250;\\
  %%CITATION = HEP-PH 0605250;%%
  U.~K\"otz, A.~Ringwald and T.~Tschentscher [APFEL Collaboration],
  % ``Production and detection of axion-like particles at the VUV-FEL: Letter of
  %intent,''
  arXiv:hep-ex/0606058;\\
  C.~Rizzo [BMV Collaboration],
  ``Laboratory and Astrophysical Tests of Vacuum Magnetism: the BMV Project,"
  2nd ILIAS-CAST-CERN Axion Training, May, 2006,
  http://cast.mppmu.mpg.de/\\
  S.~J.~Chen {\em et al.}
  %, H.~H.~Mei, W.~T.~Ni and J.~S.~Wu
  [Q \& A Collaboration],
  %``Improving ellipticity detection sensitivity for the Q \& A vacuum
  %birefringence experiment,''
  %arXiv:
  arXiv:hep-ex/0308071.
\end{thebibliography}
\end{document}